\documentclass[reprint,prl,nofootinbib,notitlepage,superscriptaddress,preprintnumbers]{revtex4-2}
\usepackage[export]{adjustbox}
\usepackage{epsfig,bm}
\usepackage{subfigure}
\usepackage[utf8]{inputenc}
\usepackage{float}
\usepackage{graphicx}
\usepackage{amsmath}
\usepackage{slashed}
\usepackage{verbatim}
\usepackage{xcolor}
\usepackage{caption}
\usepackage[colorlinks=true, linkcolor=blue, citecolor=blue, urlcolor=blue]{hyperref}
\usepackage{color}
\usepackage{soul}

\begin{document}
\preprint{MIT-CTP/6041}

\title{Short-Range Correlations Between Partons in a Proton}

\author{Jen-Chieh Peng}
\affiliation{Department of Physics, University of Illinois at
Urbana-Champaign, Urbana, IL 61801, USA}
\affiliation{Department of Physics, National Central University, Taoyuan City, 320317, Taiwan}
\email{jcpeng@illinois.edu}
\author{Krishna Rajagopal}
\affiliation{Center for Theoretical Physics, Massachusetts Institute
of Technology, Cambridge, MA 02139, USA}
\email{krishna@mit.edu}
\author{John Terry}
\affiliation{Physics Division, Argonne National Laboratory, Lemont, Illinois 60439, USA}
\email{terry@anl.gov}

\begin{abstract}
A principal lesson from recreating droplets of quark-gluon plasma (QGP) in heavy ion collisions is that it is a strongly coupled liquid, not a plasma of partons. The energy density and pressure of quarks and gluons confined within a proton are comparable to those of QGP at or just above the QCD transition temperature.
Given this similarity between protons and QGP, we propose that the investigation of correlations between nearby partons within a proton must be
a central goal for the future Electron-Ion Collider (EIC).
Here, we ask how EIC measurements can discern such short-range correlations (SRCs) of quark pairs. Doing so would 
characterize the strongly coupled
interior of a proton, 
augmenting the one-parton-at-a-time understanding of protons via (generalized) parton distribution functions, and could at the same time yield a key ingredient for the microscopic understanding of 
the liquid nature of QGP.
Motivated by the experiments that have been used to demonstrate the existence of SRCs between nucleon pairs within a nucleus, we propose using EIC observables involving measurements of a jet and a pion, together with the scattered electron,
to seek and quantify the
possible existence of SRCs between
quark pairs within a nucleon. 
The pronounced isospin dependence observed in the dominance of $np$ SRCs over $pp$ or $nn$ SRCs has played a central role in establishing the importance of SRCs among nucleons in nuclei. 
Analogously, the QCD attraction in the $ud$ diquark channel can make the $ud$ SRC stronger than the $uu$ and $dd$ SRCs, allowing a first observation of partonic SRCs.

\end{abstract}

\maketitle

\textit{Introduction and Observable}: Ultrarelativistic heavy ion collisions at the Relativistic Heavy Ion Collider (RHIC) and the Large Hadron Collider (LHC) have taught us that 
the quark-gluon plasma (QGP) that filled the universe for the first microseconds after the Big Bang is a strongly coupled liquid with lower specific viscosity than that of any other known liquid, not at all a plasma of quark and gluon quasiparticles~\cite{PHENIX:2004vcz,BRAHMS:2004adc,PHOBOS:2004zne,STAR:2005gfr,Gyulassy:2004zy,Muller:2006ee,Jacak:2012dx,Muller:2012zq,Heinz:2013th,Shuryak:2014zxa,Akiba:2015jwa,Busza:2018rrf}. As the RHIC era concludes and we look ahead toward the Electron-Ion Collider (EIC), it is natural to ask what the QCD lessons that we have learned from heavy ion collisions can teach us about what will be the big QCD questions for the EIC era. 
The EIC science program will add to our  understanding of the initial stage of heavy ion collisions~\cite{AbdulKhalek:2021gbh}, but this is not our focus here.
``Imaging'' the structure of the proton has long 
been identified as a central goal for the EIC.
In this paper, we ask how what we have learned about QGP from heavy ion collisions can teach us what to look for when the EIC focuses on the wave function of the partons in a proton.

One motivation for learning about protons from what we know about QGP comes from
noting that the first protons formed from droplets of QGP as the universe cooled through the QCD transition temperature 
$T_c\sim 155$~MeV~\cite{Borsanyi:2013bia,HotQCD:2014kol}. 
And, estimates of the entanglement entropy within a proton suggest that it is comparable in magnitude to the Gibbs entropy
of the QGP from which a proton forms
at the QCD transtion~\cite{Kharzeev:2026inq}.
Further motivation comes from comparing
the energy density within a proton to the energy density of QGP. We note that if protons were spheres with radii $r_p$ and a uniform energy density $\varepsilon_p$, namely with $\frac{4}{3}\pi r_p^3 \varepsilon_p = 938$~MeV, then if their radii
were $r_p=0.86$~fm, comparable to their charge radii,  
(or 0.49 fm, comparable to their mass radii~\cite{Kharzeev:2021qkd,Duran:2022xag}) their energy density $\varepsilon_p$ would be the same as that of QGP with a temperature $T\sim 155$~MeV (200 MeV), per lattice QCD calculations of QCD thermodynamics~\cite{Borsanyi:2013bia,HotQCD:2014kol}.
That is, the energy density within a proton is comparable to that of QGP at temperatures just above the QCD transition temperature $T_c$.
Furthermore, lattice QCD calculations of the gravitational form factors of the proton~\cite{Shanahan:2018pib,Shanahan:2018nnv,Hackett:2023rif} and electron-scattering data~\cite{Polyakov:2002yz,Burkert:2018bqq,Lorce:2018egm,Polyakov:2018zvc} suggest that to the extent that the proton has an interior pressure it too is comparable to the pressure of QGP at a temperature $T_p\sim (155-200)$~MeV, at or somewhat above $T_c$~\cite{Kharzeev:2026inq}.

In QGP with an energy density and pressure comparable to that within protons, the correlations are so strong and of such short range that 
it is impossible to define quasiparticles with well-defined mean free paths. 
Partons in QGP are not confined within hadrons, but they are very far from free as they are always strongly coupled to their neighbors. QGP is a strongly coupled fluid 
in which
the momenta of neighboring
fluid cells are strongly correlated.
These basic facts  
tell us that there must be strong SRCs between partons in QGP, but we know this only indirectly via
their collective effects on droplets 
made of thousands of partons produced in a heavy ion collision.
These basic facts also
tell us to expect
that the momenta of quarks and gluons within a proton must be strongly correlated with the momenta of their neighbors --- and in this setting we may be able to detect the SRCs directly! 
In this paper, we show  how EIC
measurements can discern strong SRCs between the momenta of partons
in a proton that are near each other in position space.

It is instructive in this regard 
to compare the study of SRCs between partons in a proton to the study of SRCs between nucleons in nuclei~\cite{Hen:2016kwk,CiofidegliAtti:2015lcu}. Evidence for nucleon-nucleon SRCs, referring to the formation of a nucleon-nucleon pair within a nucleus
with a large relative momentum between the pair and a small total pair momentum, has been found in triple coincidence
measurements of processes such as $A(p,ppn)$~\cite{Tang:2002ww}, 
$A(e,e^\prime p p)$ and $A(e,e^\prime p n)$~\cite{LabHallA:2014wqo}, 
where 
the incident proton or electron knocks 
a nucleon out of the nucleus with high-momentum transfer
and one detects the scattered beam particle, the nucleon that was knocked out, and an accompanying nucleon carrying the balance of the relative momentum of the nucleon pair. The observation of nucleon-nucleon SRCs in experiments shows that nuclei cannot be fully understood simply in terms of the single-particle shell model. The strong
short-range nucleon-nucleon force can cause a significant nucleon-nucleon 
correlation~\cite{Hen:2016kwk,CiofidegliAtti:2015lcu}. These findings lend support to a dynamic model of nuclei including ephemeral nucleon-nucleon pair fluctuations.

\begin{figure}[t]
\begin{center}
\includegraphics[height=0.43\textwidth,valign=c]{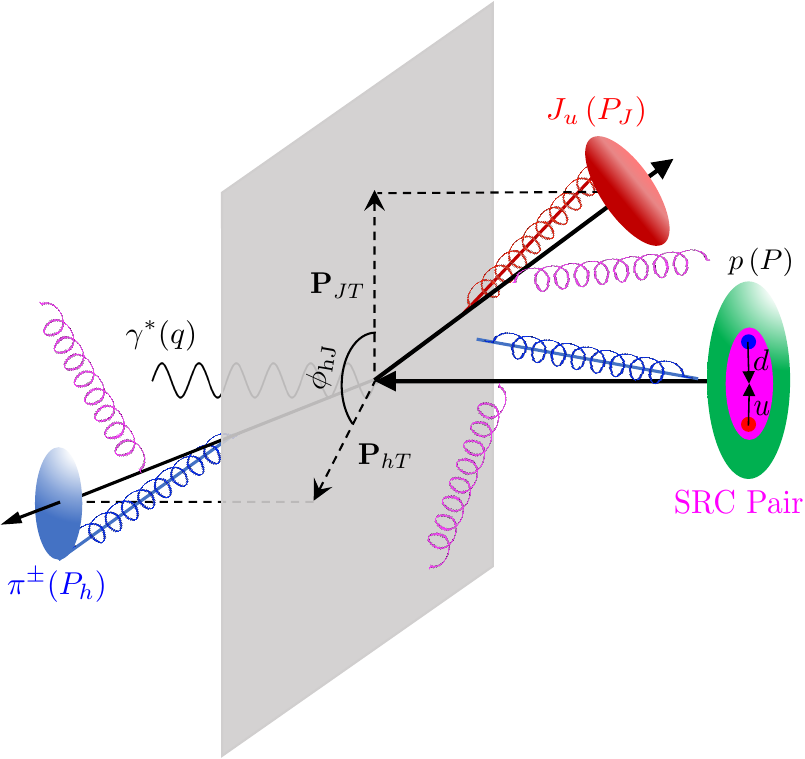}
\vspace{-0.05in}
\caption{
Schematic illustration of the observable used to probe SRCs between quark pairs in the proton in 
deep-inelastic scattering, 
shown in the Breit frame. The 
virtual photon and proton travel along the $\pm z$-axis. A $u$ 
quark from a correlated $ud$ (or $uu$) SRC pair in the proton absorbs the virtual photon 
and produces a jet with four-momentum $P_J$ in the current fragmentation 
region, while the partner $d$ (or $u$) quark emerges 
in the target fragmentation region and 
fragments into a  $\pi^-$ (or $\pi^+$) with four-momentum $P_h$. 
The transverse momenta 
$\boldsymbol{P}_{J\perp}$ and $\boldsymbol{P}_{h\perp}$ are measured 
in the $(x,y)$-plane perpendicular to the $z$-axis, and 
$\phi_{hJ}$ denotes the azimuthal angle 
between them. A $ud$ SRC produces a preference for back-to-back transverse momenta ($\phi_{hJ}\sim\pi$) between $u$-quark jets in the current region and $\pi^-$ in the target region.
}
\vspace{-0.3in}
\label{SRC_udpair}
\end{center}
\end{figure}

The EIC provides an unprecedented opportunity to look for
evidence of partonic SRCs in the proton. 
Consider the production of a jet in the current fragmentation region and a pion in the target fragmentation region,
as in Fig.~\ref{SRC_udpair}. Here, an electron
scatters off a $u$ quark which happens to be strongly correlated with a nearby $d$ quark in the wave function of the proton,
denoted SRC in Fig.~\ref{SRC_udpair}.
The scattered $u$ quark, having absorbed a large momentum from the exchanged
photon, can be detected as a jet in the current 
fragmentation kinematic region. 
The flavor of the $u$-quark jet could be 
tagged on a statistical basis by requiring
that the leading hadron in the jet is a $\pi^+$. If the struck $u$-quark came from a SRC $ud$ pair in the incident proton,
the $d$ quark from  the SRC $ud$ pair
would emerge as a $\pi^-$ in the target fragmentation  
region, shown in blue in 
Fig~\ref{SRC_udpair}. 
In an
$ep$ collider such as the EIC, the $\pi^-$ 
would be found 
near the proton beam
direction, meaning that it would have a large momentum
in the laboratory frame.
By measuring correlations
between the azimuthal angle 
$\phi$ of the $u$-quark jet
and the $\pi^-$ near the proton
beam,
the process diagrammed in Fig.~\ref{SRC_udpair} can be used to
search for correlations between the momenta of  
$u$ and $d$ quarks in the proton. 
Such SRCs among partons 
reflect strong but ephemeral pair fluctuations and do not imply the presence of parton-parton bound states.

One aspect of SRCs in nuclei 
is that they are much stronger in the $np$ channel than in the $nn$ or $pp$ channels~\cite{Tang:2002ww,Piasetzky:2006ai}.
It is also reasonable to expect that partonic $ud$ SRCs within a proton will be stronger than $uu$ or $dd$ partonic SRCs. The Pauli principle tells us that short range pairwise correlations will be strongest for pairs of partons whose wave function in color+flavor+spin is antisymmetric, and in QCD both single gluon exchange and the instanton-induced 't Hooft vertex are most strongly attractive in the channel that is separately antisymmetric in color, in flavor, and in spin.
This so-called ``good diquark'' channel has long played a role in the modeling of a baryon as a diquark and a quark~\cite{Close:1973xw,Farrar:1975yb,Lichtenberg:1982jp,Anselmino:1992vg,Isgur:1998yb,Jaffe:2003sg,Santopinto:2004hw,Cloet:2008re,Barabanov:2020jvn} and corresponding lattice QCD calculations~\cite{Alexandrou:2006cq,Francis:2021vrr,Francis:2022fdj,Francis:2022stl}, and is a principal consideration in the analysis of diquark pairing and color superconductivity in cold quark matter 
at high density~\cite{Alford:1997zt,Rapp:1997zu,Alford:1998mk,Alford:2007xm}.  
This picture is supported by covariant
Faddeev calculations, in which $ud$ pairs predominantly form compact scalar
($0^+$), isoscalar diquarks, in contrast to the more extended axial-vector
($1^+$) configurations accessible to $uu$ pairs~\cite{Cloet:2008re}.
Lattice QCD studies of diquark correlators indicate that  $ud$ pairs within a nucleon form compact configurations
with a typical size
$r_{qq} \sim 0.5\text{--}0.7~\mathrm{fm}$~\cite{Francis:2022fdj,Francis:2022stl}, suggesting a relative momentum for correlated pairs of order $|{\bf k}| \sim (300-400)~{\rm MeV}\gtrsim \Lambda_{\rm QCD}$.
All these considerations suggest that partonic SRCs in QGP and within a proton should be stronger for parton pairs that are antisymmetric in color,
in spin, and in flavor --- with the latter providing a key 
to their study at the EIC since it should be possible to differentiate $ud$ SRC from $uu$ and $dd$ SRC on a statistical basis in EIC measurements of processes like the one in Fig.~\ref{SRC_udpair}.

In the process of
Fig.~\ref{SRC_udpair}, we
consider a current jet  initiated by a struck $u$ quark 
as well as a $\pi^-$ or $\pi^+$ from the remnant of the proton.
In the absence of  SRCs, isospin symmetry implies that
the distribution of hadrons initiated by remnant $u$ and $d$ quarks are identical, meaning that the angular distributions for $\pi^-$ and $\pi^+$ 
in the target fragmentation region are identical. 
This will 
not be so if the proton features $ud$ SRCs.
We denote
the components of 
the jet and pion momenta (red and blue in Fig.~\ref{SRC_udpair}) 
perpendicular to the virtual photon momentum (perpendicular to the Breit-frame $z$-axis)
by 
$\bm{P}_{J\perp}$ and $\bm{P}_{h\perp}$.
We shall focus on $\phi_J$ and $\phi_h$, the azimuthal angle (in the Breit-frame $(x,y)$-plane) of 
$\bm{P}_{J\perp}$ and $\bm{P}_{h\perp}$,
and in particular on $\phi_{hJ}\equiv \phi_h-\phi_J$.
The presence of
$ud$ SRCs makes it more likely that 
the $u$-quark jet and a $\pi^-$ are nearly back-to-back in $\phi$, resulting in an enhancement in
the $\phi_{hJ}$ distribution around $\phi_{hJ}\sim \pi$.
$uu$ SRCs can similarly be detected by measuring a $\pi^+$ 
in the target region in coincidence with a 
$u$-quark jet in the current region.
EIC measurements can thus compare $ud$ SRCs to $uu$ SRCs within a proton by comparing the $\phi_{hJ}$ distributions for $\pi^-$ vs.~$\pi^+$ hadrons in the target fragmentation region when the jet in the current fragmentation region is tagged as a $u$-quark jet via measuring a leading $\pi^+$.

The observation of SRCs among nucleons transformed our understanding of nuclei  
and the observations of collective flow 
in heavy ion collisions  transformed our understanding of QGP. Likewise,
the experimental
observation of SRCs between partons in
a proton would have tremendous implications for our understanding of how QCD describes the strongly coupled interior of a proton,
confirming that protons cannot be fully understood in terms of (generalized) single-parton distribution functions.
We emphasize that the process  in Fig.~\ref{SRC_udpair} provides {\it direct} access to strong parton-parton SRCs,  here within a proton. 
To date, the strong, short-range, parton-parton correlations that must be present in QGP at the corresponding
energy density and pressure 
in order to explain the observed collective behavior of hundreds or thousands of partons in liquid QGP 
have not been detected directly.

\begin{figure}[t]
  \centering
  \hspace{-0.15in}\includegraphics[width=0.50\textwidth]{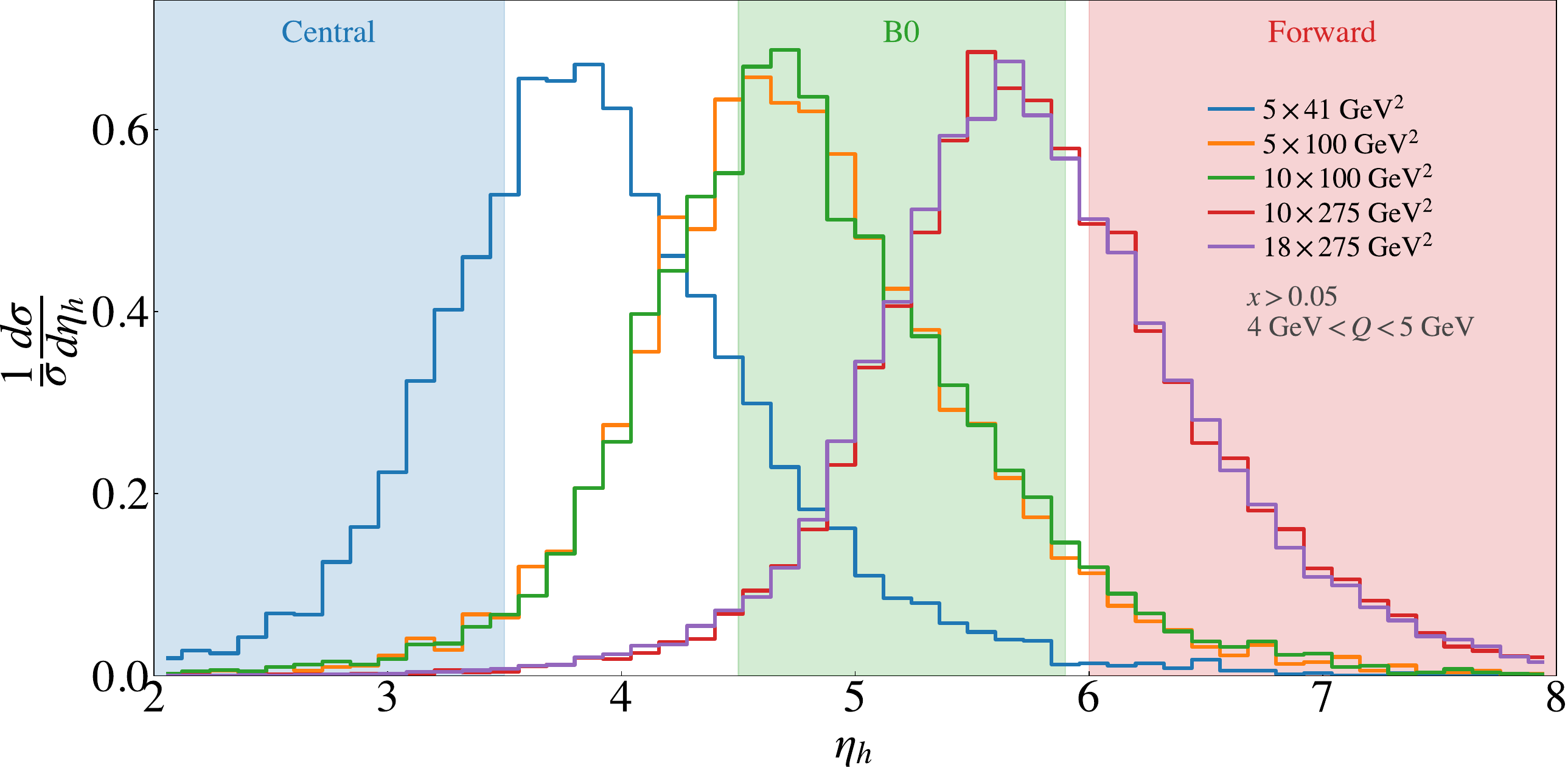}
  \vspace{-0.1in}
  \caption{
  Laboratory pseudorapidity distribution of the leading target-fragmentation
  pion at the EIC for collisions with electron$\times$proton energies of 5$\times$41 GeV$^2$, 10$\times$41 GeV$^2$, 10$\times$100 GeV$^2$, and 10$\times$275 GeV$^2$, shown together with the
  approximate acceptance of the central detector, B0, and forward detector subsystems.
  }
  \label{fig:target_eta_pT}
\end{figure}

\textit{Feasibility study at the EIC}:
The ePIC detector being developed for the EIC has been designed to measure the scattered
electron momentum and energy (and hence $Q$ and $x$) and the jet originating from the struck quark.
To assess the feasibility of the measurement that we propose, 
we must check that it can measure
$\pi^-$ and $\pi^+$ in the target fragmentation region, with  kinematics of interest.
We do so by performing a Monte Carlo study using \textsc{Pythia~8}~\cite{Sjostrand:2014zea}
for 
deep-inelastic scattering at EIC kinematics with 
various choices of electron
and proton beam energies (see 
Fig.~\ref{fig:target_eta_pT} and
Table~\ref{tab:detector_coverage})
and with kinematic cuts
\begin{align}\label{eq:kinematic-cuts}
  4~\mathrm{GeV} < Q < 5~\mathrm{GeV}, \qquad x > 0.05\ .
\end{align}
The leading target hadron is identified in the Breit frame as the most energetic hadron in the target hemisphere and is required to be a charged pion, $h=\pi^\pm$. The event is then boosted back to the laboratory frame and the pseudorapidity of the selected hadron is computed.
The resulting distributions are shown in Fig.~\ref{fig:target_eta_pT} and
quantified in Table~\ref{tab:detector_coverage}. For the $5\times 41$~GeV$^2$
configuration of beam energies, the distribution peaks at moderate forward rapidities,
$\eta_h \sim 4$, with a sizable fraction of events within the central detector
and only a small contribution in the B0 and forward regions. In
contrast, with $E_p=100$ (or 275 GeV),
the distribution is shifted to
larger pseudorapidities, peaking at $\eta_h \lesssim 5$ (or $\eta_h\lesssim 6$), and is dominated by B0
and forward rapidities, with a negligible fraction of events remaining in
the central detector.

\begin{table}[t]
\centering
\begin{tabular}{lcccccc}
\hline\hline
& & & \multicolumn{3}{c}{Event fraction ($\%$)} \\
\cline{4-6}
Subsystem & $\eta$ range & PID capability & $41$ & $100$ & $275$ \\
\hline
Central
& $|\eta|<3.5$
& ($\pi/K/p$; $e/\gamma$)
& $26.5$ & $2.9$ & $0.3$ \\
B0
& $\eta \sim 4.5$--$5.9$
& ($+/-$; $e/\gamma$)
& $16.8$ & $59.0$ & $56.5$ \\
Forward
& $\eta \gtrsim 6$
& None
& $1.0$ & $5.5$ & $34.3$ \\
\hline\hline
\end{tabular}
\caption{
Approximate kinematic coverage and measurement capabilities of the main EIC detector subsystems relevant for this analysis. The final columns show representative event fractions for three proton beam energies. The detector acceptance depends primarily on the proton beam energy, with the signal migrating toward the forward region as $E_p$ increases.
}
\label{tab:detector_coverage}
\end{table}

While geometric acceptance extends into the far-forward region, Table~\ref{tab:detector_coverage} shows that particle identification capabilities are strongly limited outside the central detector. In particular, reliably distinguishing $\pi^+$ from $\pi^-$ and from protons requires both charge tagging and hadron identification, which are only simultaneously available in the central detector. The B0-detector provides charge tagging but lacks full hadron identification, while the forward detectors provide either tracking without identification or calorimetric measurements of neutral particles.
These considerations motivate future study of the extent to which time-of-flight measurement in the B0-detector could 
distinguish $\pi^+$ from protons and motivates upgrades to ePIC
or development of a second EIC detector that improve forward acceptance and particle identification. 
For the present,
we shall assume that the measurement we propose is only feasible when the $\pi^{+/-}$ ends up in the central detector, which constrains the 
experimentally accessible phase space
for the present observable.
In light of these constraints, for the remainder of this work we 
choose beam energies $E_e\times E_p=5\times 41$~GeV$^2$ and
restrict our analysis to the region
\begin{align}\label{eq:eta-h-cuts}
  2 < \eta_h < 3.5\ .
\end{align}
This ensures that the observable is defined in a region where both charge tagging and particle identification are experimentally viable.

\textit{Modeling diquark SRCs}:
To assess the sensitivity of the proposed observable to partonic SRCs, we begin from a sample of parton-level deep-inelastic scattering events
generated with \textsc{Pythia}~8~\cite{Sjostrand:2014zea}
which provides a good baseline since we know that in its description of the proton in terms of single-parton distribution functions there are no partonic SRCs.
We select events in which, as in Fig.~\ref{SRC_udpair}, the struck quark is a \(u\) quark. In this case, the proton remnant in \textsc{Pythia} is a 
$u$ and $d$ quark 
with no correlation between their momenta
(that \textsc{Pythia} nevertheless refers to as a diquark as this describes its antitriplet color and isoscalar flavor).
This provides a controlled baseline
in which the remnant 
carries no  isospin asymmetry.

We then construct a modified event sample in which we introduce SRCs between the struck $u$ quark and the $d$ quark in the remnant by hand in 5\% or 10\% of the events,
a fraction chosen
arbitrarily,
and look for the resulting observable effects. Given the crudeness of the way we introduce SRCs, the results we shall present are not predictions for the magnitude of SRC effects at the EIC; our goal is simply to demonstrate that the observable we propose is sensitive to partonic SRCs in the proton.

To model $ud$ diquark SRCs in the incident proton, we start by
taking the $d$ quark in the remnant to have the same longitudinal momentum 
that the struck $u$ quark had before it was struck (namely a fraction $x$ of the 
momentum of the proton)
and then give the struck $u$ quark 
a momentum kick
of magnitude  $|\bm{k}| = 0.4~\mathrm{GeV}$
with a direction chosen randomly isotropically
in the frame of the $ud$ diquark
and give the $d$ quark from the remnant an equal and opposite kick.
(We have chosen the size of the kick inspired by the lattice results~\cite{Francis:2022fdj,Francis:2022stl} discussed in the Introduction.)
We apply the momentum
kicks at the parton level, prior to hadronization. 

One complication is that because of the way  \textsc{Pythia} 
implements color flow during hadronization, we cannot 
immediately do what we describe above. Instead, we must first
implement the replacement
    $(ud) \;\to\; d+ (u q) + \bar{q}$
in the remnant of the proton as described by \textsc{Pythia}, as this maintains a valid dipole color structure while allowing us to kick the $d$ quark as we wish. To avoid introducing artificial isospin asymmetry in the final state, we take the additional quark flavor  
to be $q = s$.
We partition the longitudinal momentum of the remnant among the resulting partons 
to reflect the expectation 
that the valence $u$ and $d$ quarks carry the dominant share of the 
remnant's collinear momentum, while the $s\bar{s}$ pair --- 
introduced solely as a device to maintain a valid dipole color 
structure in \textsc{Pythia} 
while allowing us to add SRCs between the struck $u$ and the $d$ from the remnant
--- carries a negligible fraction.
Concretely, we assign longitudinal momentum fractions $x$, $1-2x-.01$, 
and $0.01$ to the $d$, $(us)$,  and  
$\bar{s}$ respectively, all as fractions of the proton momentum.
Then, we can kick the struck $u$ quark in the current region and the $d$ in the target region as we have described, modeling the consequences of them having been an SRC-pair in the incident proton.
After these modifications, the partonic configuration is passed back to \textsc{Pythia} for hadronization. 
We apply the modification that we have described to  either 5\% or 10\% of the generated events, leaving the remaining events unaltered.

We then evaluate the $\phi_{hJ}$ distribution on those events in our sample  (both the unmodified sample and the samples with 5\% or  10\% of the events modified) 
that satisfy the criteria \eqref{eq:kinematic-cuts} 
and 
\eqref{eq:eta-h-cuts}, as follows.
In this model study, we use the momentum
of the struck $u$ quark
as 
a proxy for the momentum of the $u$-quark jet that an experimentalist would need to reconstruct and tag via a leading $\pi^+$.
We identify a pion in the target fragmentation region by selecting
the most energetic charged pion in the Breit frame subject to the constraint that it lies in the target hemisphere, \(p_{z}^{\rm Breit} > 0\). We further require that the pion carries a substantial fraction of the proton light-cone momentum,
    $z_{LC} \equiv q \cdot P_h/q \cdot P > 0.1$, 
thereby ensuring sensitivity to the leading remnant structure.  
In each event we can then evaluate $\phi_{hJ}$, the 
difference between the Breit-frame azimuthal angles of the pion in the target region and the struck $u$ quark.

\begin{figure}[t]
  \centering
\hspace{-0.4in}\includegraphics[width=0.53\textwidth]{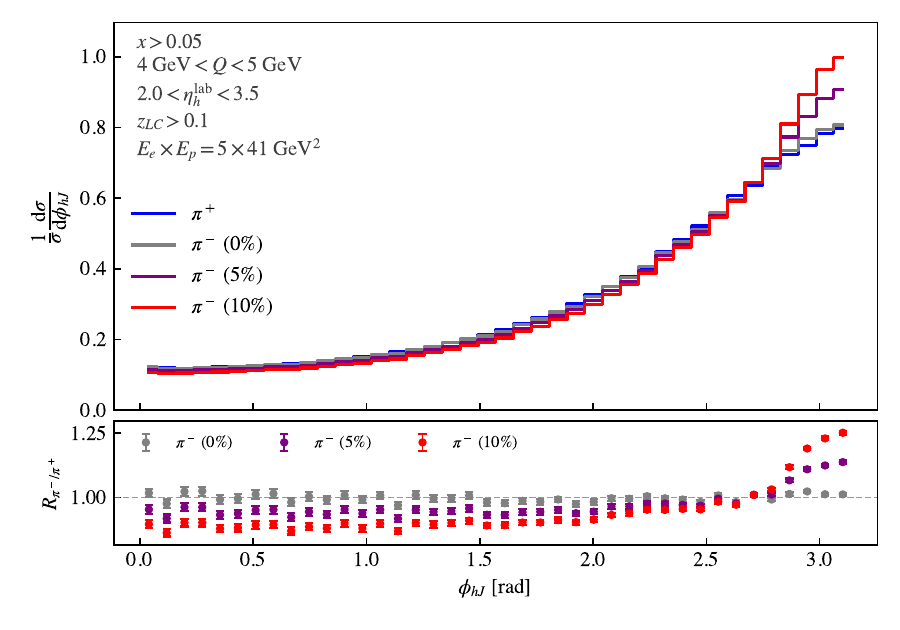}
\vspace{-0.2in}  
  \caption{Breit frame azimuthal angle correlation between the $u$-quark jet in the current  region and the hadron ($\pi^+$ or $\pi^-$) in the target region. Top panel shows the $\pi^+$ (blue) and $\pi^-$ (grey, purple or red) distributions as a function
of $\phi_{hJ}$; bottom panel shows the ratio $\pi^-/\pi^+$.
Grey: baseline \textsc{Pythia} results without modification, $\pi^-$ and $\pi^+$ distributions agree within statistics; the consequences of momentum conservation are apparent. Purple (red): results with a 5\%  (10\%) admixture of $ud$ short-range correlations in the incident proton. 
  }
\vspace{-0.2in}
  \label{fig:target_pT}
\end{figure}

The resulting azimuthal angle correlations are shown in
Fig.~\ref{fig:target_pT}. In the absence of modifications, the
distribution exhibits a clear enhancement at large angles,
$\phi_{hJ} \sim \pi$, reflecting 
transverse momentum conservation in 
the Breit frame.
The $\pi^+$ and
$\pi^-$ distributions are nearly identical, consistent with the isoscalar
nature of the remnant in the baseline sample with no SRCs. 
When either a 5\% or a 10\% admixture of events with a momentum
imbalance injected as we have described is included, modeling $ud$ SRCs in the proton, a visible distortion of the $\pi^-$ angular
correlation emerges. Enhanced $\pi^-$ correlation at $\phi_{hJ}\sim\pi$ is 
a direct manifestation of SRCs, which can be detected via the
nontrivial difference between the $\pi^+$ and $\pi^-$ $\phi_{hJ}$
distributions. 
The magnitude of the effect 
is seen in the ratio
$\pi^-/\pi^+$,  see the lower panel
of Fig.~\ref{fig:target_pT}, which deviates from unity once 
SRCs are introduced.

\textit{Closing Remarks and a Look Ahead}: Strong short-range correlations between the momenta of nearby
partons in QGP 
are manifest indirectly as near-perfect fluidity of macroscopic volumes of this form of 
matter. 
In this work, 
we have explored the possibility that strong short-range correlations 
between the momenta of nearby partons within a proton can be manifest directly in observables that can be measured at the EIC.
We have proposed an observable 
that provides a clean probe of partonic SRCs: 
the Breit-frame azimuthal angle correlation between the struck quark (manifest as a jet in the current region) and a leading hadron in the target region.

To demonstrate the sensitivity of this observable, we constructed  modified event samples in which in 5\% or 10\% of the events we gave 400~MeV momentum kicks in opposite directions to the struck $u$ quark and a $d$ quark in the proton remnant
with the same $x$. 
This serves as a crude model for the effects of 
diquark SRCs 
within a proton.  
We find that even a modest admixture of such correlated events leads to a measurable distortion of the 
correlation in $\phi_{hJ}$,
the difference in the Breit-frame azimuthal angle between the jet originating from the struck $u$ quark and the $\pi^-$ in the target region.
Because we anticipate $ud$ SRCs to be much stronger 
than $uu$ SRCs, this distortion is measurable via comparing the $\phi_{hJ}$ distribution for $h=\pi^-$ to that for $h=\pi^+$,
see 
Fig.~\ref{fig:target_pT}.
This motivates a concrete, experimentally viable, strategy for probing
partonic SRCs in the proton at the future EIC.

There are many possible extensions of this work. 
Our results in Fig.~\ref{fig:target_pT} illustrate how direct experimental consequences of partonic SRCs within the proton can be seen at the EIC; they are not calculations of SRCs and hence are not predictions for such measurements.  
Although we leave calculations of SRCs to future work, in Supplemental Materials we lay some groundwork for this
by  relating partonic SRCs to a two-quark fracture function that we introduce. 
Such future work could also treat
perturbative radiation, scale evolution and jet reconstruction
effects, all of which we have neglected  in our discussion of  jets originating from a struck $u$ quark.
Further investigation of kinematic selections and options for 
flavor tagging and jet charge measurement 
may also identify additional observables with
sensitivity to (other)
SRCs.

Although the future EIC is particularly well-suited for measuring effects of partonic SRCs within the proton, 
we strongly advocate investigating to what extent these observables can be accessed in archival HERA 
data as well as in measurements at JLAB with a somewhat lower $Q$.
In this vein,
it is interesting to note two different triple coincidence experiments that have recently been performed
using the CLAS12 spectrometer at JLAB~\cite{CLAS:2022sqt,CLAS:2026auu}. 
In the first~\cite{CLAS:2022sqt},
the scattered electron in the $ep \to e^\prime p \pi^+ X$
reaction is measured in coincidence with a
$\pi^+$ in the current fragmentation region and a proton in the target
fragmentation region
and a back-to-back correlation between the azimuthal component of
the momenta of the
two hadrons has been measured.
This result shows the
feasibility of studying the correlations between the hadrons produced in the
current and target fragmentation regions, 
and motivates repeating this measurement for both  $\pi^-$
an $\pi^+$ in the target fragmentation region.
Seeing a stronger back-to-back Breit-frame azimuthal angle correlation between
the $\pi^+$ in the current
region with a $\pi^-$ in the target region than with a $\pi^+$ could yield early evidence for partonic SRCs in the proton well before the EIC program gets underway.
This prospect motivates further simulations 
to investigate the consequences of a lower $Q$ and of measuring only a $\pi^+$ in the current region, rather than reconstructing a jet.
In the more recent CLAS12 measurement in Ref.~\cite{CLAS:2026auu},
pion-pion angular correlations have been measured in a triple coincidence experiment, albeit for a charged and a neutral pion that are both in the current fragmentation region.
Performing the measurement that we propose with a neutral pion in the target fragmentation region 
could be interesting, although further study would be needed in order to establish a baseline against which to compare such a measurement as we have done here for the observable that we have proposed.

This work demonstrates the feasibility of seeing short-range correlations between the momenta of nearby partons
in a proton.  
Experimental evidence for strong short-range correlations within the proton would be as important to our understanding of QCD as the discovery that QGP is a strongly coupled fluid. The discovery of the former would yield microscopic understanding of the latter, while transforming our image of the proton.

\acknowledgments
\textit{Acknowledgments:}
We thank Henry Klest for helpful feedback on a draft of this paper and in particular for help with understanding ePIC capabilities.
We are grateful to Or Hen, Peter Jacobs, Dima Kharzeev, Zein-Eddine Meziani, Richard Milner,  Berndt M\"uller, Duff Neill, Daniel Pablos, Dimitra Pefkou, Felix Ringer, Farid Salazar, Bill Zajc and Yong Zhao for stimulating conversations. 
This work was initiated in part at the Aspen Center for Physics, which is supported by National Science Foundation grant PHY-2210452. We gratefully acknowledge the conversations that we had there with participants in the Strongly Interacting Matter at the Electron-Ion Collider workshop. 
KR and JT are also grateful to the Kavli Institute for Theoretical Physics (KITP) for hospitality and support 
and acknowledge conversations with many participants at the Frontiers of Quark-Gluon Matter Program; this research was supported in part by grant NSF PHY-2309135 to the KITP.
Research supported in part by the U.S.~Department of Energy, Office of Science, Office of Nuclear Physics under grant Contract Numbers DE-SC0011090
and DE-AC02-06CH11357, and the NSF grant PHY-2110229.

\bibliography{refs}

\begin{widetext}
\vspace{0.01in}
\section{SUPPLEMENTAL MATERIALS}
\subsection{Fracture Function and Factorization}

In these Supplemental Materials, we relate the observable that we have introduced in this paper and partonic SRCs in the nucleon to a two-quark fracture function that we shall define.
This lays the groundwork for  calculations of partonic SRCs and for using future measurements to constrain the two-quark fracture function --- both of which are goals for future work.

Conventional fracture functions describe target-region hadron production in a regime where the partner quark in the proton remnant carries transverse momentum of order $\Lambda_{\rm QCD}$ and is unresolved at the scale of the hard process. Its degrees of freedom are integrated out and absorbed into a single-parton object $\hat{\mathcal{M}}^{\,h}_{q}$~\cite{Trentadue:1993ka,Anselmino:2011ss}. The observable studied in this work probes a different regime: the kinematics select configurations in which the partner quark carries intrinsic transverse momentum of order $k_{\rm SRC} \gtrsim \Lambda_{\rm QCD}$, so that the partner quark cannot be integrated out at leading power. The observable therefore requires an explicit resolved-partner degree
of freedom at leading power. A natural realization of this structure is
a target-collinear sector containing two resolved quark modes. In these Supplemental Materials, we develop a natural realization of this structure --- the two-quark fracture function --- and the factorized cross section that follows from the implied mode separation. The existence of a resolved partner degree of freedom follows from the
power counting discussed below. The specific operator realization and
factorized structure proposed here should be viewed as physically
motivated extensions of the conventional fracture-function framework.

\textit{Kinematics.} Working in light-cone coordinates with null vectors $n^\mu = t^\mu + z^\mu$ and $\bar n^\mu = t^\mu - z^\mu$, the momenta of the proton and jet are given by
\begin{align}
    P^\mu = \frac{Q}{x_B}\,\frac{n^\mu}{2}\,, \qquad P_J^\mu = Q\,\frac{\bar n^\mu}{2}\,,
\end{align}
where power corrections associated with the proton and jet mass are neglected. The standard DIS variables are $x_B = Q^2/(2 P \cdot q)$ and $y = (P\cdot q)/(P\cdot \ell)$. For a hadron $h$ produced in the target region we use the light-cone fraction $z_{\rm LC} = (\bar n \cdot P_h)/(\bar n \cdot P)$, related to the variable of Ref.~\cite{Trentadue:1993ka} by $z_{\rm LC} = z(1-x_B)$. Parton momentum fractions are defined as $x_i = (\bar n \cdot k_i)/(\bar n \cdot P)$, and transverse momenta $\bm{k}_{i\perp}$, $\bm{P}_{J\perp}$, $\bm{P}_{h\perp}$ are measured with respect to the Breit-frame $z$-axis. We denote  the azimuthal angle between $\bm{P}_{h\perp}$ and $\bm{P}_{J\perp}$ by $\phi_{hJ}$.

\textit{Hierarchy and resolved partner quarks.} The relevant degrees of freedom in the target-fragmentation sector are dictated by the scale hierarchy
\begin{align}
    Q \;\gg\; k_{\rm SRC} \;\sim\; P_{J\perp},\, P_{h\perp} \;\gtrsim\; \Lambda_{\rm QCD}\,,
\end{align}
with $k_{\rm SRC}$ the intrinsic 
relative  momentum associated with correlated quark pairs. At the scale of the hard process, a partner quark in a SRC carries transverse momentum parametrically above $\Lambda_{\rm QCD}$ and is therefore a resolved degree of freedom rather than a component of the soft remnant: it cannot be integrated out at leading power. The standard treatment of target fragmentation~\cite{Trentadue:1993ka,Anselmino:2011ss}, in which the partner is integrated out and absorbed into $\hat{\mathcal{M}}^{\,h}_{q}$, applies when the partner momentum is of order $\Lambda_{\rm QCD}$, but is not the appropriate description in the present regime. The leading-power target sector identified above therefore motivates a
nonperturbative object containing two target-collinear quark fields, one describing the struck quark and one describing the quark in the remnant that formed an SRC with the struck quark,
together with an identified hadron in the target fragmentation region.

A power-counting argument makes the necessity concrete. The observable is measured at $P_{h\perp} \sim P_{J\perp} \sim k_{\rm SRC} \gtrsim \Lambda_{\rm QCD}$. Generating an energetic target-region hadron with transverse momentum of this size from an unresolved (soft) remnant would require coherent buildup of transverse momentum through multiple soft interactions, a process suppressed by powers of $\Lambda_{\rm QCD}/k_{\rm SRC}$. A resolved partner quark carrying intrinsic transverse momentum of order $k_{\rm SRC}$ can naturally transfer transverse momentum of this size to hadronic final states at leading power. This is the basic EFT motivation for retaining a second collinear quark field in the operator definition.

It is useful to introduce the relative and pair momenta of the two target-collinear partons, 
\begin{align}
    \bm{k}_{\rm rel} \;\equiv\; \frac{\bm{k}_{1} - \bm{k}_{2}}{2}\,, \qquad \bm{k}_{\rm pair } \;\equiv\; \bm{k}_{1} + \bm{k}_{2}\,.
\end{align}
The momentum-space signature used to identify short-range correlations in nuclei --- a relative momentum within the pair that is large compared to the typical Fermi scale, combined with a comparatively soft pair center-of-mass momentum relative to the rest of the system --- can be translated directly into the context of this work where we are interested in
partonic short-range correlations within a nucleon as
\begin{align}
    |{\bf k}_{\rm rel}| \;\sim\; k_{\rm SRC}\,, \qquad {\bf k}_{\rm pair} | \;\sim\; \Lambda_{\rm QCD}\,.
\end{align}
The observable that we have introduced in this paper is designed to isolate a distinctive observable consequence of the analogous partonic configurations within the proton. The two-quark fracture function defined below is the operator-level object that resolves these configurations: the principal physical motivation for its definition is that our goal is to focus on SRC physics, rather than generic two-parton correlations.

The two-quark color content of a diquark within the proton decomposes as $\mathbf{3} \otimes \mathbf{3} = \bar{\mathbf{3}} \oplus \mathbf{6}$. 
In the three-valence-quark  configuration, the diquark is necessarily in the $\bar{3}$, since it must combine with the third quark to form a color singlet; beyond the valence approximation, however, higher Fock components permit the two-quark subsystem to occupy either the $\bar{3}$ or the $6$, with the overall singlet maintained by the remaining partons.
Both one-gluon exchange and
the instanton-induced 't Hooft vertex are attractive in the $\bar{\mathbf{3}}$ channel ($\mathbf{T}_1 \cdot \mathbf{T}_2 = -2/3$) and repulsive in the $\mathbf{6}$ channel ($\mathbf{T}_1 \cdot \mathbf{T}_2 = +1/3$).   Compact correlated configurations are accordingly expected to be enhanced in the $\bar{\mathbf{3}}$ channel, as supported by lattice studies of gauge-invariant quark--quark correlators~\cite{Alexandrou:2006cq,Francis:2021vrr,Francis:2022fdj,Francis:2022stl} and by covariant Faddeev calculations~\cite{Cloet:2008re} which identify scalar--isoscalar diquarks with transverse size $r_{qq} \sim 0.5$--$0.7~\mathrm{fm}$. The position--momentum uncertainty relation then implies $k_{\rm SRC} \gtrsim \Lambda_{\rm QCD}$ for these configurations. The color projection itself does not define an SRC: a generic two-quark fracture function contains both correlated and uncorrelated configurations across both color channels. SRC contributions are discriminated primarily by transverse momentum, with the $\bar{\mathbf{3}}$ channel projection used here as an operator-level choice that focuses on the attractive channel where correlated configurations are expected to be much more important.

\textit{Two-quark fracture function.} The leading-power target sector identified above requires a nonperturbative object containing two target-collinear quark fields together with an identified hadron in the target fragmentation region. A natural gauge-invariant realization, built from SCET collinear quark building blocks $\chi_{\bar n}$ and projected onto the color $\bar{\mathbf{3}}$ channel, is
\begin{align}
    \mathcal{M}^{\,h}_{qq'/P}&\!\left(x_1, x_2, z_{\rm LC}, \bm{k}_{1\perp}, \bm{k}_{2\perp}, \bm{P}_{h\perp}\right)
    \equiv \frac{1}{4z_{\rm LC}\, N_c(N_c-1)}\,
    \left(\frac{\slashed{n}}{2}\right)_{\!\alpha\beta}\!
    \left(\frac{\slashed{n}}{2}\right)_{\!\gamma\delta}
    \nonumber\\
    &\times \sum_X \int \frac{d(\bar n \cdot \xi_1)\, d^2\bm{\xi}_{1\perp}}{(2\pi)^3}\,
    \frac{d(\bar n \cdot \xi_2)\, d^2\bm{\xi}_{2\perp}}{(2\pi)^3}\,
    e^{i x_1\, P\cdot \xi_1 + i x_2\, P\cdot \xi_2 - i \bm{k}_{1\perp}\cdot \bm{\xi}_{1\perp} - i \bm{k}_{2\perp}\cdot \bm{\xi}_{2\perp}}
    \nonumber\\
    &\times \left\langle P \left|\, \epsilon^{ace}\,\bar\chi^{\,a}_{\bar n,q,\alpha}(\xi_1)\, \bar\chi^{\,c}_{\bar n,q',\gamma}(\xi_2)\right| h, X\right\rangle \left\langle h, X \left|\, \epsilon_{bde}\,\chi^{\,b}_{\bar n,q,\beta}(0)\, \chi^{\,d}_{\bar n,q',\delta}(0) \right| P\right\rangle\,,
    \label{eq:Mdefn}
\end{align}
with $\xi_i = (0,\, \bar n\cdot \xi_i,\, \bm{\xi}_{i\perp})$. The Dirac projectors $\slashed{n}/2$ acting on the two bilinears select the unpolarized leading-twist component. Recall that $x_1$, $x_2$ and $z_{\rm LC}$ specify the light-cone momenta of the struck quark, the partner quark from the proton remnant, and the measured pion in the target fragmentation region, all as fractions of the proton momentum.
The light-cone fractions satisfy $x_1 + x_2 \leq 1$, ensuring that the spectator system has positive light-cone momentum. Gauge invariance requires appropriate Wilson lines connecting the quark fields; collinear Wilson lines are absorbed into the $\chi_{\bar n}$ building blocks, and soft radiation associated with the target remnant is expected to be encoded in the gauge-invariant operator definition. The precise Wilson-line structure and  factorization properties of the operator 
$\mathcal{M}$ introduced in Eq.~\eqref{eq:Mdefn}
remain to be established in a future operator-level analysis.

The object defined above is not a double parton distribution. The two partons in a double parton distribution each participate in distinct hard interactions, and the nonperturbative input describes their joint distribution prior to scattering. Here, only the struck quark $q$ enters the hard scattering; the partner quark $q'$ remains associated with the target fragmentation region and is correlated with the identified hadron $h$. The natural interpretation is therefore a generalized fracture function in which the partner is retained as a resolved degree of freedom, rather than a double parton distribution.

Summing over the partner flavor and integrating over its momentum recovers the conventional single-parton fracture function introduced in Refs.~\cite{Trentadue:1993ka,Anselmino:2011ss}:
\begin{align}
    \sum_{q'} \int dx_2\, d^2 \bm{k}_{2\perp}\, \mathcal{M}^{\,h}_{qq'/P} \;\longrightarrow\; \hat{\mathcal{M}}^{\,h}_q\,.
\end{align}
This is a matching statement: the standard single-parton object emerges once the partner-quark degrees of freedom are integrated out. Further integrating degrees of freedom by summing over hadrons and integrating $z_{\rm LC}$ reproduces the Trentadue--Veneziano momentum sum rule, $\sum_h \int_0^{1-x_B} dz_{\rm LC}\, z_{\rm LC}\, \hat{M}_q^h(x_B,z_{\rm LC}) = (1-x_B)\, f_1^q(x_B)$, relating the collinear marginalized object to the standard quark PDF.

At tree level, the struck-quark momentum is fixed by the partonic process, $x_1 = x_B$ and $\bm{k}_{1\perp} = \bm{P}_{J\perp}$. The spectator momentum is then integrated:
\begin{align}
    \mathcal{M}^{\,h}_{qq'/P}\!\left(x_B, z_{\rm LC}, \bm{P}_{J\perp}, \bm{P}_{h\perp}\right) \;\equiv\; \int dx_2\, d^2 \bm{k}_{2\perp}\, \mathcal{M}^{\,h}_{qq'/P}\!\left(x_B, x_2, z_{\rm LC}, \bm{P}_{J\perp}, \bm{k}_{2\perp}, \bm{P}_{h\perp}\right)\,,
\end{align}
where the same symbol $\mathcal{M}^{\,h}_{qq'/P}$ denotes the integrated and unintegrated forms, distinguished by their argument lists.

\textit{Factorization.} The hierarchy $Q \gg k_{\rm SRC} \sim P_{J\perp}, P_{h\perp} \gtrsim \Lambda_{\rm QCD}$ separates the dynamics into three sectors: a hard scattering at the scale $Q$, a current jet collimated along $\bar n^\mu$ at virtuality $P_{J\perp}^2$, and a target-fragmentation sector containing the two resolved target-collinear quarks and the identified hadron at the scale $k_{\rm SRC}$. The leading-power cross-section follows the structure implied by these sectors, namely
\begin{align}
    \frac{d\sigma}{d\mathcal{PS}\, d\phi_{hJ}}
    &= \sigma_0\, \sum_{q, q'} e_q^2\, H_q(Q)\, J_q(Q) \int d^2 \bm{P}_{J\perp}\, d^2 \bm{P}_{h\perp}\;
    \delta\left(\phi_{hJ} - \frac{\bm{P}_{h\perp}\cdot \bm{P}_{J\perp}}{\left|\bm{P}_{h\perp}\right| \left|\bm{P}_{J\perp}\right|} \right)
    \nonumber\\
    &\quad\times \mathcal{M}^{\,h}_{qq'/P}\!\left(x_B, z_{\rm LC}, \bm{P}_{J\perp}, \bm{P}_{h\perp}\right)\,,
\end{align}
where $d\mathcal{PS} \equiv dx_B\, dy\, dz_{\rm LC}$, $\sigma_0$ is the inclusive DIS Born prefactor, $H_q(Q)$ is the hard matching coefficient, and $J_q$ is the standard exclusive quark jet function. The hard function depends only on the struck-quark flavor (through $e_q^2$ and perturbative matching), while all dependence on the partner-quark flavor $q'$ resides in the two-quark fracture function. We propose this structure as the natural leading-power realization of the identified mode separation; a complete operator-level proof, including the precise Wilson-line and soft-mode structure, is reserved for future work.

The mode separation also organizes soft radiation in a natural way. Soft modes associated with the outgoing struck quark are absorbed into the jet function $J_q$, in parallel with the standard inclusive-jet treatment. Soft modes involving the incoming struck quark, the incoming partner quark, and the target-side remnant are absorbed into the gauge-invariant definition of $\mathcal{M}^{\,h}_{qq'/P}$, which should therefore be viewed as a \emph{dressed} target-fragmentation object rather than a bare two-quark correlator. The detailed implementation of this organization --- the precise Wilson-line structure, the rapidity factorization, and the evolution of the two-quark fracture function --- remains to be established in a future operator-level analysis. We emphasize that the open questions concern this implementation, not the existence of the leading-power sectors themselves: the resolved partner quark is a robust consequence of the hierarchy of scales inherent in the physics of SRCs within a proton, the observable that we have introduced in this paper, and the related two-quark fracture function.

\textit{The observable.} The ratio plotted in the lower panel of Fig.~3,
\begin{align}
    R_{\pi^-/\pi^+}(\phi_{hJ}) = \frac{d\sigma_{\pi^-}/d\phi_{hJ}}{d\sigma_{\pi^+}/d\phi_{hJ}}\,,
\label{eq:RatioDefn}
\end{align}
where the cross-sections are differential in $\phi_{hJ}$, is sensitive to the flavor-tagged content of the two-quark fracture function and hence to $ud$ SRCs as we have described in this paper. In Fig.~3, the cross-sections are integrated over $P_{J\perp}$ and $P_{h\perp}$, as well as over $y$ and $z_{\rm LC}$ subject to the kinematic cuts described in the main text; in principle, one could instead bin in $(x_B, y, z_{\rm LC}, P_{J\perp}, P_{h\perp})$ to access more differential information. In the absence of any SRCs (or if, contrary to the expectations that we have described in this paper, the $ud$, $uu$ and $dd$ SRCs were equally strong) the ratio~\eqref{eq:RatioDefn}
would
approach unity up to residual charge-dependent 
remnant
effects unrelated to SRCs. Generic remnant effects --- soft fragmentation differences between $\pi^+$ and $\pi^-$, charge-dependent hadronization from an unresolved remnant --- may modify the charge composition in the ratio, 
but such effects are expected to act primarily on the collinear fragmentation and to be largely confined to low transverse momentum. 
The characteristic signature expected from resolved $ud$ SRC configurations is the combination of charge dependence and back-to-back ($\phi_{hJ}$ around $\pi$)
enhancement with $P_{h\perp}$ and  $P_{J\perp}$ larger than $\Lambda_{\rm QCD}$. 
The localization of the deviation of 
$R_{\pi^-/\pi^+}(\phi_{hJ})$ away from unity
in the back-to-back, $P_\perp \gtrsim \Lambda_{\rm QCD}$, region --- together with its presence in the ratio, where ordinary remnant backgrounds are expected to largely 
cancel --- distinguishes the SRC signal from charge-dependent fragmentation backgrounds.
Measurements of $R_{\pi^-/\pi^+}(\phi_{hJ})$ across the full kinematic range in $(x_B, z_{\rm LC}, P_{h\perp}, P_{J\perp})$
would provide sufficient handles to use experimental data to constrain the two-quark fracture function
and, ultimately, to confront it with lattice and Faddeev-equation predictions for two-quark correlations.  In the longer term, measurements with different flavor-tagged current jets, 
measurements of other 
final state hadrons in the target fragmentation region, as well as
measurements of complementary observables such as $\Lambda$ polarization in the target region,
should be investigated
as they may yield analogous probes of
quark-gluon and quark-antiquark 
short-range correlations and  corresponding generalizations of the two-quark fracture function that we have introduced.
\clearpage
\end{widetext}

\end{document}